\documentclass[conference]{IEEEtran}
\usepackage[utf8]{inputenc}
\usepackage{cite}
\usepackage{amsmath,amssymb,amsfonts,flushend,tabularx,subcaption}
\usepackage{algorithmic}
\usepackage{graphicx}
\usepackage{textcomp}
\usepackage{xcolor}
\usepackage{verbatim}
\usepackage{float}

\def\BibTeX{{\rm B\kern-.05em{\sc i\kern-.025em b}\kern-.08em
    T\kern-.1667em\lower.7ex\hbox{E}\kern-.125emX}}

\newcommand{\bparagraph}[1]{\par\noindent\textbf{#1}}
\newcommand{\smallsection}[1]{\bparagraph{\textit{#1}}\hfill\\}
\newcommand{\highlight}[1]{{\color{red} {#1}}}

\begin{document}
\title{\textsc{The Naked Sun:} Malicious Cooperation\\Between Benign-Looking Processes}

\author{
    \IEEEauthorblockN{
        Fabio De Gaspari\IEEEauthorrefmark{1},
        Dorjan Hitaj\IEEEauthorrefmark{1},
        Giulio Pagnotta\IEEEauthorrefmark{1}, 
        Lorenzo De Carli\IEEEauthorrefmark{2} and
        Luigi V. Mancini\IEEEauthorrefmark{1}
    }
    \IEEEauthorblockA{
    \IEEEauthorrefmark{1}
        Dipartimento di Informatica\\
        Sapienza Universit\`{a} di Roma, Italy\\
        Email: \{degaspari, hitaj.d, pagnotta, mancini\}@di.uniroma1.it
    }\\
    \IEEEauthorblockA{\IEEEauthorrefmark{2}
        Department of Computer Science\\
        Worcester Polytechnic Institute\\
        Email: ldecarli@wpi.edu
    }
}

\maketitle

\thispagestyle{plain}
\pagestyle{plain}

\begin{abstract}
  Recent progress in machine learning has generated promising results in behavioral malware detection. Behavioral modeling identifies malicious processes via features derived by their runtime
  behavior. Behavioral features hold great promise as they are intrinsically related to the functioning of each malware, and are therefore considered difficult to evade. Indeed, while a significant amount of results exists on evasion of static malware features, evasion of dynamic features has seen limited work.

  This paper thoroughly examines the robustness of behavioral malware detectors to evasion, focusing particularly on anti-ransomware evasion. We choose ransomware as its behavior tends to differ significantly from that of benign processes, making it a low-hanging fruit for behavioral detection (and a difficult candidate for evasion). Our analysis identifies a set of novel attacks that distribute the overall malware workload across a small set of cooperating processes to avoid the generation of significant behavioral features.  Our most effective attack decreases the accuracy of a state-of-the-art classifier from 98.6\% to 0\% using only 18 cooperating processes. Furthermore, we show our attacks to be effective against commercial ransomware detectors even in a black-box setting.

\end{abstract}
\section{Introduction}
\label{sec:Introduction}

The problem of automatic malware detection is a difficult one, with no full solution in sight despite decades of research. The traditional approach---based on analysis of static signatures of the malware binary (e.g., hashes)---is increasingly rendered ineffective by polymorphism and the widespread availability of program obfuscation tools~\cite{4413008, OKane:2011:OHM:2051779.2052030}. Using such tools, malware creators can quickly generate thousands of binary variants of functionally identical samples, effectively circumventing signature-based approaches.

As a result, in recent years the focus of the community has increasingly shifted towards dynamic, behavior-based analysis techniques. Behavioral approaches sidestep the challenges of obfuscated binary analysis. Instead, they focus on the runtime behavior of malware processes, which is difficult to alter without breaking core functionality, and is therefore considered a reliable fingerprint for malware presence. This strong push towards malware behavioral analysis, coupled with the recent improvements in the field of Machine Learning (ML), has resulted in a multitude of ML-based behavioral approaches to malware detection. Popular techniques range from modeling of system call sequences~\cite{5665796} to full-fledged, fine-grained modeling of process behavior~\cite{continella_shieldfs:_2016, mehnaz_rwguard}.
At first sight, the ML-based approach appears to hold great promise: ML applications to cybersecurity have been successful in many fields, and malware detection is no exception. In general, the behavior of malware differs significantly from that of benign processes, and ML techniques allow to differentiate between them with extremely high accuracy. Moreover, ML-based approaches are also able to correctly classify unseen malware samples, as long as they are trained on an appropriate training set and the new samples exhibit
some form of anomalous behavior with respect to benign processes. 

Despite the incredible success of ML, a growing body of work has cast a shadow over its robustness in adversarial settings. Research has shown that ML classifiers in a wide range of security applications are vulnerable to attacks, such as adversarial examples~\cite{Goodfellow2014ExplainingAH} and evasion~\cite{biggio_evasion:2013}. Our work assesses the robustness of recently-proposed ML applications to behavioral ransomware detection~\cite{continella_shieldfs:_2016, mehnaz_rwguard}. We use ransomware as case study due to both the gravity of the threat (e.g.,~\cite{ransomware_atlanta:2018,ransomware_uk:2017}), and the fact that---given its highly distinctive behavioral profile---ransomware is a nearly ideal target for ML-based detection. Our results show that \textbf{it is possible to craft ransomware that accomplishes the goal of encrypting all user files, and at the same time avoids generating any significant behavioral features}. Our newly-discovered attacks have fairly low implementation complexity, do not limit ransomware functionality in any significant way, and were found to be effective against a set of academic and commercial anti-ransomware solutions. Moreover, our attacks are successful even in a black-box setting, with no prior knowledge of the the model internals, its training data, or the features used by the ML model.
The core of our approach is an algorithm that cleverly distributes the desired set of malware operations across a small set of cooperating processes\footnote{In Isaac Asimov's 1957 science fiction novel \textit{The Naked Sun}, a crime is committed by a group of robots. Each robot performs an apparently innocuous action that complies with the First Law of Robotics (A robot may not injure a human being or, through inaction, allow a human being to come to harm), however the combination of such actions results in murder.}. While our work has focused on obfuscating ransomware-related features, the underlying principles are general and likely to apply to a wide range of ML detectors that analyze the runtime behavior of different types of malware.

\bparagraph{Our contributions:}
\begin{itemize}
      \item We perform a comprehensive analysis of characteristic features typically used to detect ransomware, and define techniques and criteria for evasion.
      \item We assess the robustness of current state-of-the-art behavioral ransomware detectors, showing how it is possible to design ransomware that completely evades detection. In particular, we propose three novel evasion techniques: \textit{process splitting}, \textit{functional splitting}, and \textit{mimicry}.
      \item We implement and evaluate Cerberus, a proof-of-concept prototype of a ransomware following our approach, proving that our evasion technique is practical.
      \item We evaluate our novel evasion techniques against multiple state-of-the-art ML detectors, as well as against a leading commercial behavioral detector. Results show that our techniques are effective and successfully evade detection, even in a black-box setting.
      \item We evaluate the dependence of our attack on the dataset used. Results show that our evasion techniques are effective even without access to the dataset used to train the target classifiers. 
    \item We discuss the applicability of our evasion techniques to other types of malware, and potential countermeasures. 
   
\end{itemize}

The remainder of the paper is structured as follows: Section~\ref{sec:Background} provides background on adversarial ML and ransomware detection. Section~\ref{sec:Attacks} describes our novel evasion techniques. Section~\ref{sec:Features} analyzes features used for ransomware detection, and discusses general principles for evading detection. Section~\ref{sec:Implementation} describes a proof-of-concept ransomware implementing our approach, while Section~\ref{sec:Evaluation} evaluates it against a suite of state-of-the-art detection techniques. Section~\ref{sec:Discussion} discuss applications of our techniques beyond ransomware, and possible countermeasures.  Section~\ref{sec:RelatedWork} reviews related work, Section~\ref{sec:Ethical} discusses the ethical implications of our work, and Section~\ref{sec:Conclusions} concludes the paper.

\section{Background}
\label{sec:Background}

\subsection{Adversarial ML}
The core problem of adversarial ML can be stated in a simplified form as follows. Consider a multi-dimensional feature space $V$, where a vector $v \in V$ may represent properties (features) of an underlying object of interest $o$. For example, a process may be represented by a vector encoding the frequencies of certain system calls. Furthermore, consider a classification function $f : V \rightarrow C$ mapping vectors in $V$  to classes in a given set $C = {c_1,...,c_n}$. Typically the goal of an adversarial ML attack is to cause $f$ to misclassify one or more feature vectors, i.e. forcing the classifier to make mistakes in mapping objects to labels. Many classifiers used in security distinguish between a benign and a malicious object class, i.e., $C = \{B, M\}$. In this settings, the problem therefore becomes---given a vector $v$ belonging to class $M$---to cause $f(v) = B$. One way to ``trick'' $f$ is to alter its training dataset, causing it to learn an incorrect boundary between classes. This constitutes a poisoning attack~\cite{Biggio:2014:PBM:2666652.2666666}. Conversely, \textit{adversarial sample generation} consists in picking a victim classifier which has been trained on a correctly-labeled dataset, and crafting one or more malicious objects in such a way that they get classified as benign. In our work we focus on the latter attack.

The literature proposes several techniques to generate adversarial feature vectors directly from a formal representation of a target classifier (e.g.,~\cite{papernot2016limitations, kantchelian2016evasion}). These techniques typically are agnostic to the type of object being studied, and work purely in feature space. There is then a second line of work which investigates the complementary problem, given a successful adversarial feature vector, to generate a corresponding object (e.g., given the feature vector that a stealth malware should exhibit not to be detected, generate the actual binary of the malware). Past work demonstrates use of adversarial ML to generate stealth PDF exploits~\cite{xu2016automatically, Srndic:2014:PEL:2650286.2650798, biggio_evasion:2013}, Android malware~\cite{Yang:2017:MDA:3134600.3134642, demontis2017yes, 10.1007/978-3-319-66399-9_4}, Flash malware~\cite{maiorca2017adversarial}, and a variety of other dangerous objects. Our work falls within the realm of adversarial sample generation. In the rest of this paper, we demonstrate (i) heuristics to generate adversarial feature vectors for ransomware behavior, and (ii) a proof-of-concept ransomware prototype whose behavior generates the target adversarial feature values.

\subsection{Behavioral Ransomware Detection}
\label{sec:BehavioralRansomwareDetection}

The literature presents several recent works on ransomware detection based on behavioral features~\cite{continella_shieldfs:_2016, mehnaz_rwguard, kirda_redemption, amin_kharraz_unveil:_2016, scaife_cryptolock_2016}. UNVEIL~\cite{amin_kharraz_unveil:_2016} and its successor Redemption~\cite{kirda_redemption} detect suspicious activity by computing a score using an heuristic function over various behavioral features: file entropy changes, writes that cover extended portions of a file, file deletion, processes writing to a large number of user files, processes writing to files of different types, back-to-back writes. Similarly, CryptoDrop~\cite{scaife_cryptolock_2016} maintains a ``reputation score''---indicating the trustworthiness of a process---computed based on three main indicators: file type changes, similarity between original and written content, and entropy measurement. Additionally, CryptoDrop also uses deletion and file type funneling (reading/writing a small set of file types) as secondary indicators. 

Our attack is motivated by a review of all the approaches cited above; for evaluation however we selected two of them, which are described in greater detail in the following. The selection was based on practical considerations: both approaches were published in highly visible venues, and in both cases the authors kindly provided enough material (code and/or datasets) and support to enable us to run their software. Our evaluation also includes a commercial product from \textbf{[anonymized]} (discussed at the end of this section).

\begin{figure}[t]
  \includegraphics[width=\columnwidth]{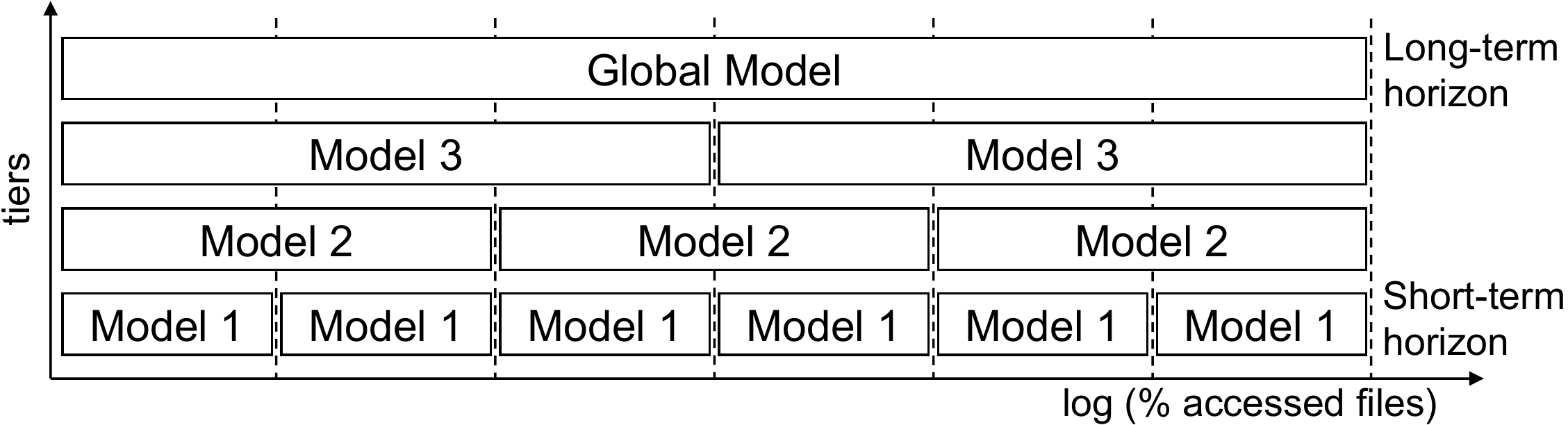}
  \vspace{-0.25in}
  \caption{Incremental models in ShieldFS (reproduced from~\cite{continella_shieldfs:_2016})}
  \vspace{-0.2in}
  \label{fig:shieldfs_structure}
\end{figure}

\subsubsection{ShieldFS}
ShieldFS~\cite{continella_shieldfs:_2016} is a technique for identifying ransomware processes at file-system level and transparently roll back file changes performed by processes deemed malicious. Ransomware detection is based on ML models of well- and ill-behaved processes. Detection is performed at the process level by using a hierarchy of random forest classifiers tuned at different temporal resolutions. Using different temporal resolutions allows ShieldFS to take into account both short- and long-term process history when performing classification, which is crucial to detect code injection-based ransomware. 
ShieldFS uses features typically associated with ransomware operation for the classifier, such as \#folder-listing operations, \#read operations, \#write operations, \#rename operations, percentage of file accessed among all those with same extension and average entropy of data passed to write operations.

ShieldFS divides the lifetime of each process in up to 28 \textit{ticks}; ticks do not represent fixed interval of times; instead, they define fractions of the overall set of files accessed by a process. Ticks are exponentially spaced; the first tick is reached when a process has accessed 0.1\% of the files on the filesystem; the last when a process has accessed 100\% of the files. Whenever a certain tick $i$ is reached, ShieldFS computes the features over multiple intervals. The first interval covers operations between ticks $i-1$ and $i$. Each of the remaining intervals ends at tick $i$ and begins further in the past compared to the previous one. Features computed over each interval are fed to a dedicated model for classification. Figure~\ref{fig:shieldfs_structure} (reproduced from~\cite{continella_shieldfs:_2016}) shows the first six ticks in the lifetime of a process, and the various intervals covered by each model. A process is considered malicious if positively detected for $K=3$ consecutive ticks.

ShieldFS also employs a system-wide classifier that computes and classifies feature values across all processes in the system. This classifier is however only used to disambiguate ambiguous results from per-process classifiers. In other words, if a per-process model over a certain interval cannot determine whether the process is malicious or not, system-wide features are computed over the same interval and fed to the system-wide classifier. When our attack is successful, individual processes are always classified as benign with high confidence and therefore the system-wide classifier is never triggered.

\subsubsection{RWGuard}

RWGuard~\cite{mehnaz_rwguard}, by Mehnaz et al., is a ransomware detector which leverages multiple techniques: process behavior, suspicious file changes, use of OS encryption libraries, and changes to decoy files. We do not discuss decoy and library-based detection as it is orthogonal to our work. Compared to ShieldFS, RWGuard uses a relatively simple detector consisting of a random forest classifier that analyzes process behavior using a 3 seconds sliding window. The features used by the classifier include the number of various low-level disk operations performed by each process under analysis. The behavioral classifier is complemented by a file monitor component which computes four metrics after each write operation: a similarity score based on similarity-preserving hashing, size difference, file type change and file entropy. Significant changes in any of first three metrics and/or high file entropy are interpreted as a sign of ransomware activity.

The detection process of RWGuard consists of three steps: when the behavioral classifier detects a suspicious process activity, the file monitor component is invoked to validate the initial detection. If both modules agree that the activity is suspicious, a third module, the File Classification module, is invoked to assess if the encryption operation is benign or malicious. Only after all three modules agree on the maliciousness of the suspicion activity, then the responsible process is considered malicious. When our attack is successful, individual processes are classified as benign by the behavioral module, and the remaining modules are not invoked.

\subsubsection{\textbf{[anonymized]}}
Several commercial anti-ransomware solutions exist; for our work, we chose to evaluate \textbf{[anonymized]}. 
\textbf{[anonymized]} does not provide details on the inner workings of their product; the company however states that their product leverages machine learning.
These indications suggest some type of behavioral classifier. For our evaluation, we use version \textbf{[version]}.

\section{Evading ML Detectors}
\label{sec:Attacks}
\begin{figure*}[h!]
\centering
  \includegraphics[width=0.9\textwidth]{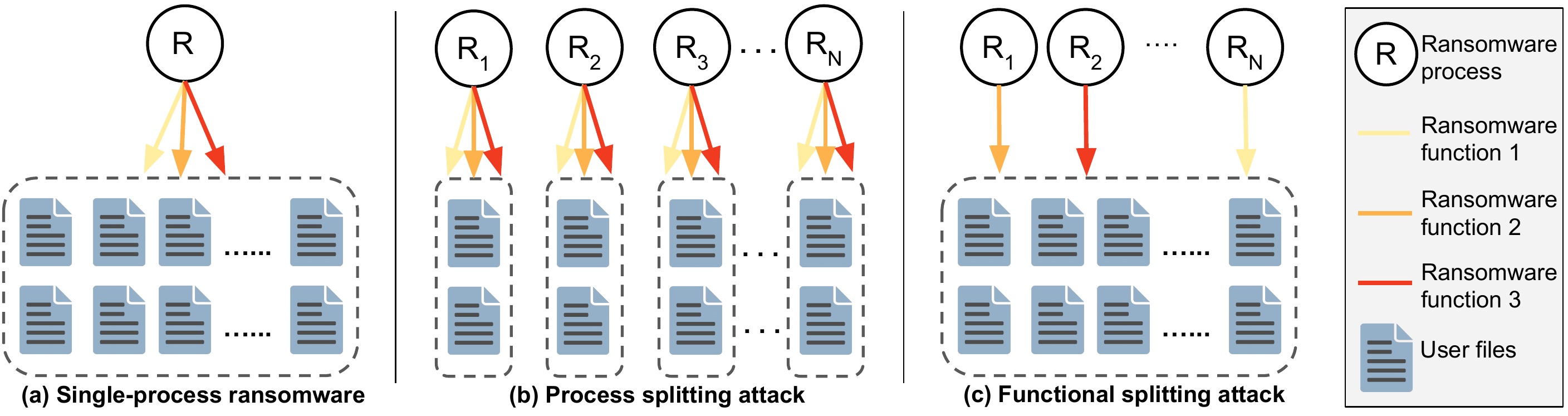}
  \vspace{-0.1in}
  \caption{Process and functional splitting attacks}
  \vspace{-0.1in}
  \label{fig:splittings}
\end{figure*}
\subsection{Process Splitting}
\label{sec:BinarySplitting}
Behavioral classifiers are designed to use features that are considered inextricably linked with malicious behavior and generally not present in benign applications. Our approach is based on the insight that behavioral detectors collect these features on a per-process basis to model the behavior of a given application. For instance, ransomware detectors profile processes based on features such as entropy of write operations or number of read/write/directory listing operations. We exploit this limitation by devising a novel evasion technique based on distributing the malware operations across multiple independent processes:
each process individually appears to have a benign behavior. However, the aggregated action of all these processes results in the intended malicious malware behavior. It is important to note that this is not just a limitation of current behavioral classifiers, but it is rather an inherent restriction of process behavioral modeling, as there is no straightforward way to identify a set of processes working independently to achieve a common final goal. However, coordinating $N$ processes requires communication among them, which could be used to identify the set of cooperating processes. To avoid this, covert communication techniques can be used. Section~\ref{sec:Implementation} discusses inter-process communication and how it is currently implemented in our prototype. The remainder of this section describes our three proposed approaches to evade classification, in increasing order of complexity: process splitting, functional splitting and mimicry.

\emph{Process Splitting} is the simplest and most straightforward multi-process ransomware evasion approach. In process splitting, the ransomware behavior is distributed over $N$ processes, each performing $1/N$ of the total ransomware operations. Effectively, this approach implements a form of data parallelism: each individual process performs all the ransomware operations on a subset of the user files. The intuition is that ransomware classifiers are trained on traditional, single-process ransomware, which exhibits extremely high number of operations such as directory listing, read and write.
Splitting the ransomware over $N$ independent processes allows to reduce the number of such operations performed by each individual processes, since each process only encrypts a subset of all files. If we split the original ransomware enough times, the number of operations performed by each individual \emph{process-split ransomware process} becomes low enough that the classifier is unable to detect the ransomware.

While this technique is simple, our experiments show it can be extremely effective even against complex classifiers (see Section~\ref{sec:Evaluation}). Moreover, the approach can be easily tailored against specific ML models by following a simple 
process splitting procedure, as illustrated in Figure~\ref{fig:splittings}(b).
Given a trained classifier we want to evade, we run our ransomware and query the model to check if it is detected. If the ransomware is detected, we further split its operations and query the model again. We continue splitting the ransomware operations over an increasing number of processes and querying the classifier until the desired evasion rate is achieved.

\subsection{Functional Splitting}
\label{sec:FuncSplitting}
While process splitting is very effective in reducing the accuracy of ransomware classifiers, completely evading detection can be challenging. Indeed, depending on the robustness of the target classifier, process splitting might require creating a very large number of processes. This process explosion may be used to detect the presence of ransomware.

A more well-rounded approach to classifier evasion is \emph{Functional Splitting}, as illustrated in Figure~\ref{fig:splittings}(c). Ransomware processes perform a set of operations (or \emph{functions}) to encrypt user files, such as reading, writing or directory listing. When using functional splitting, we separate each of these ransomware functions in a process group: each process within the group (called \emph{functional split ransomware process}) performs only that specific ransomware function. Within each group, we can further apply the process splitting technique to reduce the number of the specific operation performed by an individual functional split ransomware process. The intuition behind the functional splitting approach is that ML classifiers use groups of features to classify processes. If a process only exhibits a small subset of the features that the model associates to ransomware, then it will not be classified as malicious. Functional splitting takes this concept to the extreme, having each functional split ransomware process only exhibit a single ransomware feature.

\subsection{Mimicry}
\label{sec:Mimicry}
Functional splitting is extremely effective against current state-of-the-art ransomware classifiers. Moreover, it does not suffer from the process explosion issue that affects process splitting. However, it could be feasible to train an ML model to recognize this particular evasion attack. Typical benign processes perform several different types of functions, therefore an ML model could be trained to differentiate between benign processes and functional split ransomware processes. 

To avoid this potential drawback, we propose a third evasion attack: \emph{Mimicry}. Rather than splitting ransomware processes into individual functional groups, each ransomware process is designed to have the same functional behavior as a benign process, effectively making it indistinguishable from other benign applications. The intuition behind the mimicry approach is that behavioral ML models classify samples based on the expression of a given set of features. Ransomware processes exhibit some characteristic features, while different benign applications exhibit different sets of features to different degrees. By splitting ransomware into multiple processes---and having each individual process exhibit only features displayed by benign processes---it becomes impossible for a classifier to distinguish between the runtime behavior of \emph{mimicry ransomware processes} and benign processes. Effectively, mimicry ransomware processes are modeled after benign processes and exhibit only features that benign processes exhibit. Moreover, the degree to which each feature is exhibited by each mimicry process (e.g., how many read/write operations are performed) is kept consistent with that of benign processes. 

The end result of the mimicry approach are ransomware processes that act exactly like benign processes. However, the collective behavior of all the mimicry processes results in the desired malicious end goal. Section~\ref{sec:Features} discusses which features are characteristic of ransomware processes, and how we can limit the occurrence of each of these features in order to mimic the behavior of benign processes.

\section{Features Discussion}
\label{sec:Features}
Behavioral classifiers exploit the marked behavioral differences between benign programs and malware in order to detect malicious samples. In the context of ransomware, such classifiers rely on a wide range of features that all ransomware programs must exhibit in order to reach their goal. This section discusses these features and analyze their robustness to evasion. Many of the features described here are also displayed by benign processes, and each feature by itself does not provide strong evidence for or against ransomware behavior. However, when considered together, these features highlight a very unique program behavior proper of ransomware processes. 

\subsection{Write Entropy}
\label{sec:WriteEntropy}
The end goal of ransomware is to encrypt users' files and collect a ransom payment in exchange for the decryption key. Typical encrypted data is a pseudorandom string with no structure, and exhibits maximum entropy~\cite{mehnaz_rwguard}, while structured data written by benign programs is assumed to have considerably lower entropy. Consequently, entropy of write operations appears to be a useful feature to differentiate ransomware from benign processes. In fact all state-of-the-art ML ransomware detectors use entropy of write operations as a feature, in one form or another~\cite{kirda_redemption,mehnaz_rwguard,continella_shieldfs:_2016,kirda_unveil}.

\subsubsection{Evasion}
Entropy as a feature for ransomware detection can be used at different levels of granularity: (1) overall file entropy~\cite{mehnaz_rwguard}, (2) average read-write operations difference~\cite{kirda_redemption,kirda_unveil}, and (3) individual write operations~\cite{continella_shieldfs:_2016}.
Feature (1) does not allow accurate differentiation between ransomware and benign processes, as nowadays 

most common file types are compressed for efficiency, including file types generally targeted by ransomware such as pdf, docx, xlsx, video and image files. Consequently, the overall file entropy for this file types, as measured by current state-of-the-art approaches relying on Shannon entropy~\cite{Lin1991}, is comparable to that of an encrypted file. For what concerns feature (2), in our research we analyzed several file types with their associated programs, and found out that in general benign processes working on compressed formats exhibit numerous very high entropy read and write operations.

It is worth pointing out, however, that despite the considerations above our dataset also shows a non-negligible difference in the average entropy of individual file \textit{write operations}. Such averages are $0.4825$ for benign processes vs $0.88$ for ransomware, with range $[0-1]$. Despite this somewhat counter-intuitive result, it is still straightforward to evade feature (3). Average write entropy can be skewed simply by introducing artificial, low-entropy write operations that lower the average write entropy for a ransomware process, bringing it in line with that of benign processes.

\subsection{File Overwrite}
Different ransomware families use different techniques to encrypt user files. However, one feature that is common across ransomware families is that the original file is fully overwritten, either with the encrypted data or with random data to perform a secure delete~\cite{kirda_unveil}. On the other hand, benign processes rarely overwrite user files completely. Therefore, file overwrite is a valuable feature that can be exploited to classify between ransomware and benign processes.

\subsubsection{Evasion}
In order to evade this feature, we need to limit the percentage of a file overwritten by a single ransomware process. Maintaining this percentage within the range exhibited by benign processes can be easily achieved with our proposed multi-process ransomware approach. It is sufficient to distribute write operations to a given file over multiple ransomware processes, each of which only overwrites a portion of the file. Each individual process does not show any suspicious behavior, but the aggregated action of all the processes still overwrites the whole file.
\vspace{-0.1in}
\subsection{Directory Traversal}
In order to maximize the amount of damage for the victim, ransomware typically encrypts every file in any given user directory. Therefore, ransomware processes issue \emph{open} operations for every file in a directory. This behavior is fairly distinctive and not common in benign processes, except for some particular cases. Therefore, when coupled with other meaningful features, directory traversal can serve as a useful indicator to detect a ransomware process.

\subsubsection{Evasion}
The directory traversal feature can be easily evaded using our multi-process ransomware approach. Indeed, by setting an upper bound on the number of files each ransomware process can access in a given directory, we can heavily reduce the expression of this feature.
\vspace{-0.1in}
\subsection{Directory Listing}
In order to encrypt user files, ransomware first need to discover all files of interest. In order to perform file discovery, ransomware issue a very large amount of file listing operations. While some types of benign processes also exhibit similar behavior (e.g., Windows Explorer), this feature can be a valuable indicator of ransomware activity when coupled with other meaningful features.

\subsubsection{Evasion}
The directory listing feature can be easily evaded using our multi-process ransomware approach. Indeed, having multiple processes allows us to distribute the listing operations and therefore limit the incidence of this feature for each individual process.
\vspace{-0.1in}
\subsection{Cross-File Type Access}
Typically, benign processes only access a fixed subset of file types (e.g., a pdf viewer will access pdf files, but not docx). On the other hand, a ransomware process will access all important user files, regardless of the coherence of their file types, in order to encrypt them (e.g., pdf, xlsx, jpeg). This cross-file type access behavior can therefore serve as a useful feature to detect ransomware operations when coupled with other meaningful features.

\subsubsection{Evasion}
Cross-file type access can be evaded in a similar fashion to the directory traversal feature. It is sufficient to separate ransomware processes into different groups, and have each process in a group access only a coherent subset of file types. This is enough to make a ransomware process indistinguishable from benign processes from the point of view of file type access.

\subsection{Read/Write/Open/Create/Close Operations}
Ransomware accesses and encrypts as many files as possible on the victim's directories to maximize the damage and ensure the payment of a ransom. This behavior results in an abnormally large amount of file operations such as \emph{read, write, open, close} and, for some ransomware families, \emph{create}. Typical benign processes rarely access so many files in a single run, except for some particular cases (e.g., files indexer).

\subsubsection{Evasion}
Our multi-process approach allows to evade this feature. By using multiple coordinated processes to encrypt user files, each individual process only needs to access a subset of all user files. By varying the number of ransomware processes used, we can limit how many file operations each individual ransomware process performs.

\subsection{Temp Files}
Some ransomware families use temporary files as buffer to encrypt files before overwriting them, or as a temporary data storage while copying or removing the original files~\cite{mehnaz_rwguard}. While several benign programs also generate temporary files, this behavior can be useful when used together with other indicators to detect the presence of ransomware~\cite{mehnaz_rwguard}.

\subsubsection{Evasion}
Ransomware does not necessarily need to use temporary files to encrypt user data. Therefore, the most obvious and effective way of evading this feature is not creating temporary files. Regardless, our multi-process technique can be applied to keep the number of .tmp files generated by each individual process in line with the average number of .tmp files used by benign applications.

\subsection{File Type Coverage}
For the purpose of ransomware, not all files are equally important. Some file types are far more likely to contain important data for the victim than others (e.g., a docx file vs a cfg). Ransomware strive to access and encrypt all files with relevant extensions~\cite{continella_shieldfs:_2016}, such as all docx files in a directory tree, in order to maximize damage. On the other hand, benign programs typically only access a fraction of these files. 

\subsubsection{Evasion}
While it can be tricky for traditional ransomware to evade this feature, it is fairly straightforward when using our proposed approach. A natural consequence of distributing ransomware over multiple processes is that
each individual process accesses only a fraction of the total user files.
To evade the file type coverage feature, it is sufficient to make sure that each ransomware process only accesses and encrypts a portion of all the files of a given type.

\subsection{File Similarity}
Encrypting a file completely changes its content, since the original data is overwritten with pseudorandom data. Typical benign processes rarely alter a file in a way that its content is overall completely different from the previous version. On the other hand, ransomware always completely change the whole content of a file when encrypting it. Therefore, 
overall file similarity before and after write operations from a given process is a strong feature to detect ransomware operation~\cite{scaife_cryptolock_2016}.

\subsubsection{Evasion}
Traditional ransomware always changes the whole file, either by encrypting it or by performing a secure delete operation. Therefore, there does not appear to be a simple way to avoid file similarity feature. However,
our approach can evade this feature with a technique like that proposed
for the file overwrite feature. By having each ransomware process encrypt only a portion of any given user file, we preserve the overall file similarity after each individual write operation, and no individual process changes the whole file content.

\subsection{File-Type Change}
Files are structured data, and most file types are characterized by some \emph{magic bytes}~\cite{magic_bytes} within the file content. When a file is fully encrypted, the magic bytes are also altered, effectively changing the file type signature. Benign program operations do not generally alter file types. On the other hand, ransomware operation always do. Therefore, file type change can be a useful feature to detect ransomware operation.

\subsubsection{Evasion}
The file type change feature can be easily evaded, even by traditional ransomware. Indeed, it is enough for the ransomware to preserve the original magic bytes of the file. Even if more complex file-type techniques are used, such as looking for changes in the overall structure of a given file, a ransomware can still evade this feature by maintaining the overall file structure during encryption with techniques such as format-preserving encryption~\cite{10.1007/978-3-642-05445-7_19}.

\subsection{Access Frequency}
In order to maximize the damage to the victim and minimize the time window to intervene and stop the attack, ransomware aims to encrypt user files as quickly as possible. To do so, a typical ransomware performs write operations on different files in short time windows. Some ransomware classifiers therefore use the write access frequency of a process as a feature to differentiate between benign and malicious processes~\cite{kirda_redemption}.

\subsubsection{Evasion}
Our analysis indicates that even benign processes often exhibit high file-write access frequency. Analyzing the ShieldFS process dataset~\cite{continella_shieldfs:_2016}, we found that it is not uncommon for benign processes to perform write operations to different files in short time intervals, making this feature rather weak for ransomware classification.
Moreover, using our multi-process approach allows to decrease the ransomware access frequency to be in line with that of any benign processes, while maintaining a high file encryption throughput due to parallelization.

\begin{table}
\centering
\begin{tabularx}{\columnwidth}{|l|X|}  \hline
  DL & Directory listing operation  \\ \hline
  RD & Read operation               \\ \hline
  WT & Write operation              \\ \hline
  RN & Rename operation             \\ \hline
  OP & Open operation               \\ \hline
  CL & Close operation              \\ \hline
  FRD & Fast read operation         \\ \hline
  FWT & Fast write operation        \\ \hline
  FOP & Fast open operation         \\ \hline
  FCL & Fast close operation        \\ \hline
  \{X,Y\} & Functional group of processes performing op. X and Y \\ \hline
\end{tabularx}

\caption{Notation}
\label{tab:notation_table}
\vspace{-0.2in}
\end{table}

\subsection{Other Features}
This section covered what we found to be the most used and robust features employed by current ransomware classifiers. Other features were proposed in the literature to improve detection accuracy. However, we found that these are either very weak (e.g., file size change~\cite{mehnaz_rwguard}), or extremely similar to other features that we already discussed (e.g., file type funneling~\cite{scaife_cryptolock_2016}). Therefore, evasion is either not necessary or achieved in a similar way to what discussed. 

\section{Implementation}
\label{sec:Implementation}
This section presents the implementation of Cerberus, our ransomware prototype implementing the mimicry evasion technique, as well as our re-implementation of the ShieldFS classifier.

\subsection{The Cerberus Prototype}
\label{sec:Cerberus}
Splitting ransomware functionality over independent processes requires coordination. Ensuring that each ransomware process only expresses features typical of a benign process further complicates the implementation.

 Here, we briefly describe Cerberus, a new ransomware prototype developed to demonstrate the feasibility of our evasion techniques. The Cerberus prototype implements both the \textit{functional splitting} and \textit{mimicry} attacks. Functional splitting separates ransomware functions in different \emph{functional groups}: a process in any given group performs only the specific ransomware functions assigned to that group. For instance, ransomware processes in to the read-write functional group only perform read and write operations. Cerberus implements functional splitting by separating ransomware operations in three groups: (1) directory list, (2) write and (3) read-rename. Read and rename are performed in the same functional group mainly for implementation convenience. Note that separating read and write does not require additional open operations, as Cerberus uses the Windows API {\tt \small DuplicateHandle()} to share handles to opened file between collaborating ransomware processes. We could have considered additional features for the implementation of functional splitting, as discussed in Section~\ref{sec:Features}. However, since the goal of Cerberus is merely to prove the feasibility of our evasion techniques, we considered only the most important features exhibited by every ransomware family. Section~\ref{sec:CerberusEvaluation} shows that the features considered are enough to evade even commercial ransomware detectors in a black-box settings.

\begin{table}[t]
\small
\centering
\begin{tabular}{|c|c|c|c|r|}  \hline
 \textbf{DL}    & \textbf{RD}   & \textbf{WT}   & \textbf{RN}   & \textbf{\% of Processes}    \\\hline 
  \checkmark    & \checkmark    & \checkmark    & \checkmark    & 19.07        \\\hline 
  \checkmark    & \checkmark    & -             & -             & 18.37        \\\hline
  -             & \checkmark    & \checkmark    & \checkmark    & 16.35        \\\hline
  -             & \checkmark    & -             & -             & 11.44        \\\hline
  \checkmark    & \checkmark    & \checkmark    & -             & 7.60        \\\hline 
  -             & \checkmark    & -             & \checkmark    & 6.85        \\\hline 
  -             & -             & -             & \checkmark    & 6.21        \\\hline 
  -             & \checkmark    & \checkmark    & -             & 5.61        \\\hline 
  \checkmark    & -             & -             & -             & 3.55        \\\hline 
  -             & -             & \checkmark    & \checkmark    & 2.18        \\\hline
  \checkmark    & \checkmark    & -             & \checkmark    & 1.76        \\\hline 
  -             & -             & \checkmark    & -             & 0.42        \\\hline 
  \checkmark    & -             & -             & \checkmark    & 0.38        \\\hline 
  \checkmark    & -             & \checkmark    & -             & 0.13        \\\hline 
  \checkmark    & -             & \checkmark    & \checkmark    & 0.08        \\\hline 
  
\end{tabular}
\vspace{0.1em}
\caption{Behavioral profiles exhibited by benign processes and their presence in the dataset.}
\label{tab:benign_process_behaviour}
\vspace{-0.2in}
\end{table}

To implement the mimicry attack in Cerberus, we performed a statistical analysis on the behavior of benign processes from the ShieldFS dataset, which contains traces from 2245 unique benign applications collected over a month. Table~\ref{tab:benign_process_behaviour} shows that we can identify
a few
behavioral classes that represent most benign processes in the dataset. For our implementation, we chose the 2nd and 3rd most represented classes: directory listing-read and read-write-rename. We chose these because they are highly represented in the dataset of benign processes, as well as because no ransomware process belongs to any of these two classes (all the 383 real ransomware in our dataset exhibits directory listing, read, write and rename operations together). Within each class, we strive to maintain the same ratio between operations as exhibited by benign processes. To achieve this, we introduce dummy operations, such as null reads or empty writes, to maintain the exact operation ratio of benign processes.

\subsection{Ransomware Interprocess Communication}
In order to orchestrate individual processes and achieve the ransomware's goal, it is necessary to coordinate them properly, which requires some form of inter-process communication. For instance, writer processes need to know the original content of the file to encrypt, which is provided by reader processes. However, standard inter-process communication mechanisms such as pipes generate additional I/O Request Packets (IRP)~\cite{irp_microsoft}, which are used by current state-of-the-art detectors to calculate features. Therefore, using standard inter-process communication would skew the behavior of the mimicry ransomware processes, potentially making them easier to detect. The Cerberus prototype implements a stealthy communication technique based on direct memory access, which allows us to avoid generating additional IRP traces for process communication. In particular, we leverage a feature of the Windows API that allows processes belonging to the same user to read/write directly to each other's address space, without the need for any special permissions or memory sharing. This is possible since all ransomware processes are started by the same user. The only requirement is for the ransomware processes to know the address range from the address space of the process they wish to read/write to, which in our case is passed as an argument during process creation. This is mostly for implementation convenience, as we could easily use more sophisticated covert channel techniques to share the memory address in a stealthy manner.

It should be noted that an improved detector could, for instance, use lower level system call hooks to intercept direct memory communication. However, there are many alternative approaches ransomware processes can employ to communicate in a stealth manner, from lower level kernel functions to covert channel techniques. Even ignoring covert channels and assuming a detector can monitor all possible types of process communication, how to do so without introducing undue overhead is not straightforward. Given that Cerberus is merely an experimental prototype, we did not implement more complex forms of stealth communication. We further discuss the issue in Section~\ref{sec:Discussion}.

\subsection{ShieldFS}
\label{sec:ShieldfsImplementation}
As we could not obtain the original code or a prototype for ShieldFS due to patenting issues, we re-implemented the ShieldFS classifier exactly as described, interacting with the ShieldFS's authors to clarify any potential misunderstanding. We split the ShiedFS dataset in training and testing set following a $10:1$ ratio and trained each of the ShieldFS model's tiers with the appropriate feature vectors from benign and ransomware traces. As in the original paper, we trained multiple classifiers for each of the 28 tiers, covering percentage of file accessed from $0.1\%$ up to $100\%$. Each classifier is implemented as a random forest of 100 trees. We validated our implementation on the training set and obtained results in line with the original classifier. More specifically we performed a one-machine-off cross validation, as done in the original paper, where benign data from one machine is selectively removed from the training set and used for the testing. The average accuracy of our classifier came out at 98.6\%, which is slightly higher that the $97.7\%$ overall performance reported in~\cite{continella_shieldfs:_2016}.
\section{Evaluation}
\label{sec:Evaluation}
This section presents the experimental evaluation of our evasion techniques. In particular, we aim at answering the following research questions: (1) \emph{is our theoretical attack technique effective in avoiding detection?} In Section~\ref{sec:LogEvaluation}, we apply our techniques to traces generated executing traditional ransomware, and show that process splitting, functional splitting and mimicry effectively avoid detection;
(2) \emph{can our theoretical attack evade detection when implemented in a real-world setting?} In Section~\ref{sec:CerberusEvaluation} we evaluate our prototype Cerberus in a virtual machine, showing that our theoretical attacks can be implemented and are effective in the real world; (3) \emph{do our evasion techniques generalize, evading classifiers trained on different datasets?} In Section~\ref{sec:CerberusEvaluation} we show that the mimicry attack, modeled on the ShieldFS dataset and implemented in Cerberus, successfully evades detection of RWGuard, which is trained on a different dataset; (4) \emph{is our attack effective in a black-box setting against commercial behavioral ransomware detectors?} Section~\ref{sec:CommercialDetectorEval} shows that Cerberus successfully evades detection of \textbf{[anonymized]}, a leading commercial behavioral ransomware detection tool.

\subsection{Dataset and Experimental Setup}

\begin{table}
  \small
  \centering
  \begin{tabular}{|l|c|c|}
    \hline
    \textbf{Type} & \textbf{Benign} &\textbf{Ransomware} \\\hline
    \#Unique Applications & 2245  & 383   \\\hline
    \#Applications Training Set     & 2074  & 341   \\\hline
    \#Applications Testing Set      & 171   & 42    \\\hline
    \#IRPs [Millions]   & 1763  & 663.6 \\\hline
  \end{tabular}
  \caption{Dataset details}
  \label{tab:sfs_dataset}
  \vspace{-0.1in}
\end{table}
  
Our trace-based evaluation leverages a dataset provided to us by the authors of ShieldFS. Table~\ref{tab:sfs_dataset} summarizes this dataset; further details can be found in~\cite{continella_shieldfs:_2016}. To train our classifiers, we divided the data on benign processes from the 11 machines comprising the dataset into: $10$ machines for the training set and one for the testing set. For the $383$ ransomware samples, which include different ransomware families, we use $341$ for training and $42$ for testing.

In order to test our Cerberus prototype, we created a realistic virtual machine-based testbed, consisting of a VirtualBox-based Windows-10 VM. We based the VM user directory structure and file types on the disk image of an actual office user. File contents were extracted from our own machines and replicated as needed. In total, our VM is comprised of $33625$ files for a total of $\sim 10GB$, distributed over $150$ folders.

\subsection{Trace-Based Evaluation}
\label{sec:LogEvaluation}
This section presents the trace-based evaluation of process splitting, functional splitting and mimicry attacks.
This evaluation uses the I/O Request Packets (IRP) traces~\cite{irp_microsoft} of real ransomware from the ShieldFS dataset.
The testing dataset contains $42$ unique ransomware samples, which include different ransomware families. For each ransomware, the IRP trace contains the complete list of I/O operations performed by the ransomware process. Both ShieldFS and RWGuard extract the ransomware features used for detection, such as number of read/write operations, directly from the IRP Traces. In this section, our evaluation is based on the ShieldFS dataset~\cite{continella_shieldfs:_2016}.

Our evaluation simulates multiple processes by splitting the IRP trace of a single ransomware in multiple traces, based on the specific evasion technique under evaluation. Successively, we compute the feature vector for each individual trace as if it were an individual ransomware process. Finally, we query the classifier and compute the percentage of feature vectors classified as belonging to a ransomware. Table~\ref{tab:notation_table} introduces the notations that we will use in the remainder of this section.

    \begin{figure}[t]
        \centering
        \includegraphics[width=.9\columnwidth]{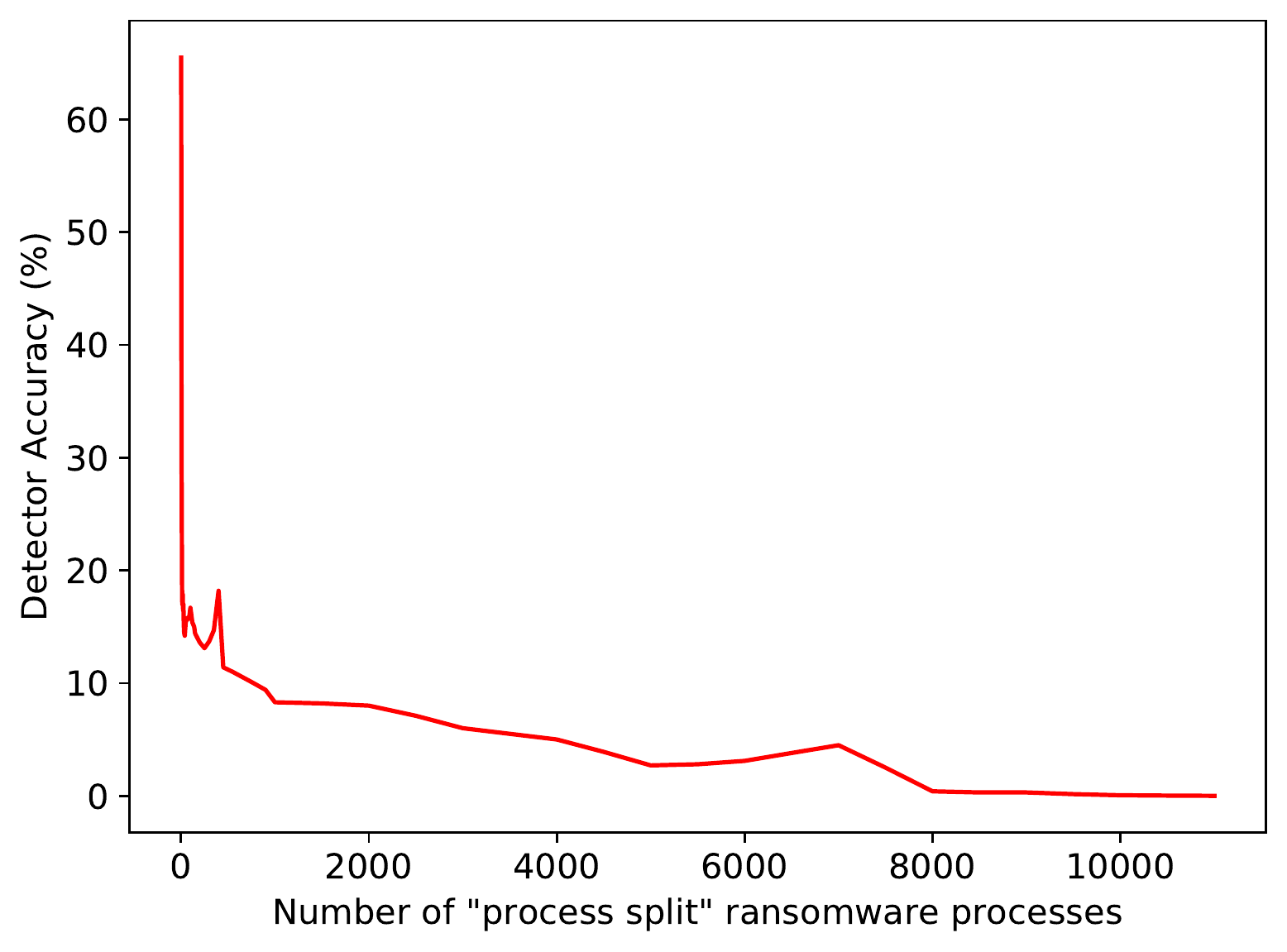}
        \caption{Evaluation of process splitting against ShieldFS}
        \label{fig:binary_splitting_eval_sfs}
        \vspace{-0.2in}
    \end{figure}

\smallsection{1) ShieldFS}
This section evaluates the effectiveness of our techniques against the ShieldFS ransomware detector.

\bparagraph{- Process Splitting:} 
As mentioned in Section~\ref{sec:BinarySplitting}, process splitting evenly splits the operations performed by a ransomware process over $N$ processes. In a process-split ransomware, all processes exhibit almost identical behavior and characteristics. We begin our evaluation by splitting the original ransomware trace in multiple traces, querying the classifier in each trace. We increase the number of traces
until complete evasion is achieved. We evaluate process splitting with $42$ unique ransomware traces, which include different ransomware families. We compute the feature vector for each process-split ransomware, query the classifier and compute the percentage of feature vectors flagged as malicious. Figure~\ref{fig:binary_splitting_eval_sfs} illustrates our results. 
We can see that ShieldFS accuracy decreases already after a single split, going from single-process $98.6\%$ accuracy down to $65.5\%$ on a two-process ransomware. Further splitting incurs diminishing returns: going from two to ten processes results in $20.1\%$ accuracy ($45.4\%$ decrease), and going from ten to one hundred processes results in $16.69\%$ accuracy ($3.41\%$ decrease). Completely evading the ShieldFS classifier requires approximately $11000$ processes. The requirement of such a large number of processes to achieve full evasion is a clear drawback of this simplistic approach. It is reasonable to imagine a countermeasure that can detect process-split ransomware by monitoring the process creation behavior at a system-level. Large swaths of newly created processes that exhibit similar behavior can then be clustered and analyzed as if they were a single process.

\bparagraph{- Functional Splitting:}
As discussed in Section~\ref{sec:Features}, ransomware exhibits several distinctive features. 
Functional splitting (see Section~\ref{sec:FuncSplitting}) exploits the reliance of current behavioral classifiers on the presence of most of these features to detect ransomware. This section evaluates the effectiveness of evasion based on the lack of expression of certain features against ShieldFS. The ShieldFS classifier is trained on six features: \#folder listing (DL), \#file reads (RD), \#file write (WT), \#file rename (RN), file type coverage and write entropy. Our evaluation focuses on the four main operations performed by ransomware---DL, RD, WT, RN---and split ransomware processes based on these four functional groups. Finally, we assess how each functional split ransomware process performs against the detector. Note that focusing only on these 4 features makes it harder to evade the detector, since we make no attempt to evade the remaining 2 features.

 \begin{figure*}[t]
	    \centering
	    \begin{subfigure}{.9\columnwidth}
	        \centering
                \includegraphics[width=.9\columnwidth]{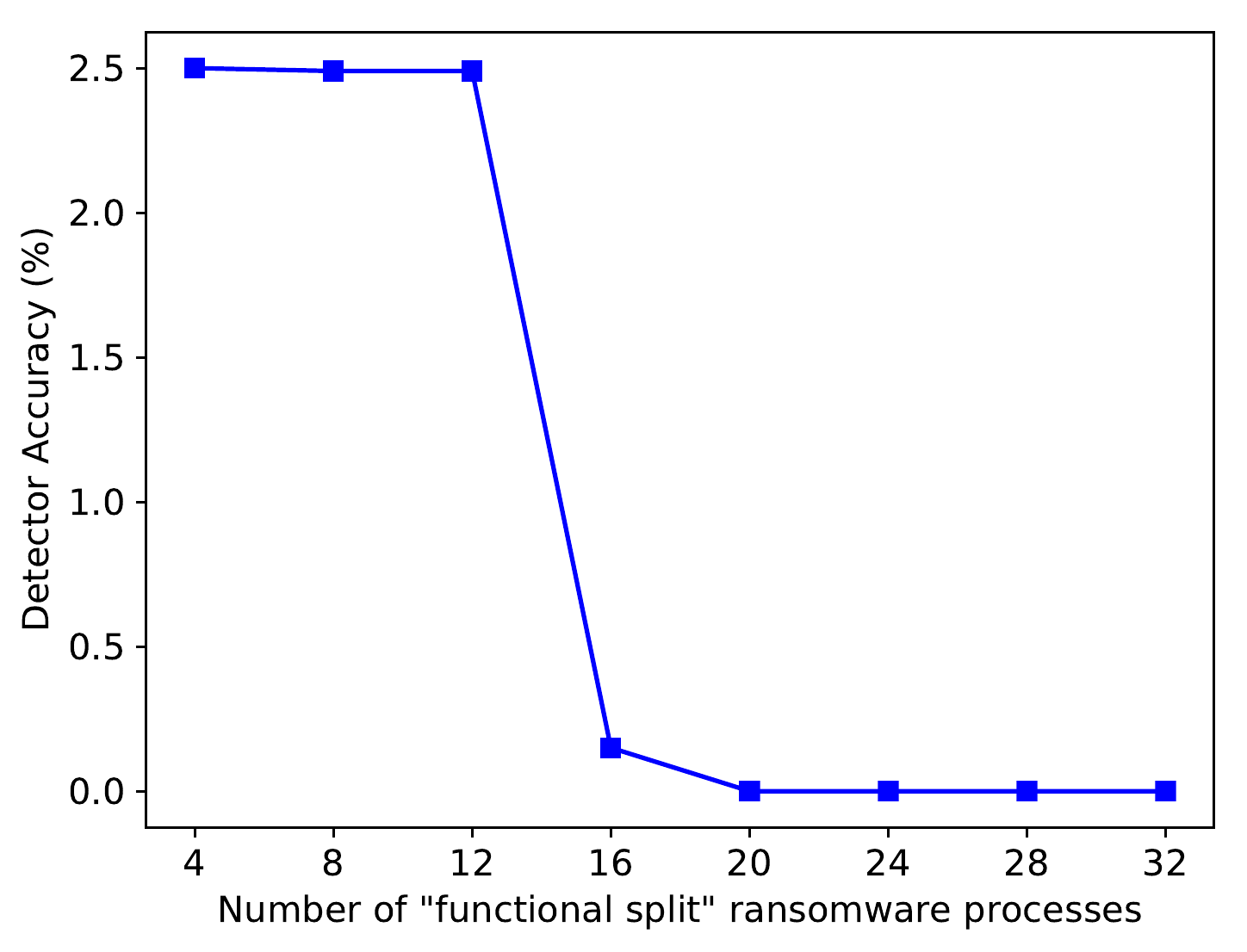}
                \vspace{-0.1in}
            \caption{Single functional splitting}
            \label{fig:single_func_split}
        \end{subfigure}
        \begin{subfigure}{.9\columnwidth}
            \centering
            \includegraphics[width=.9\columnwidth]{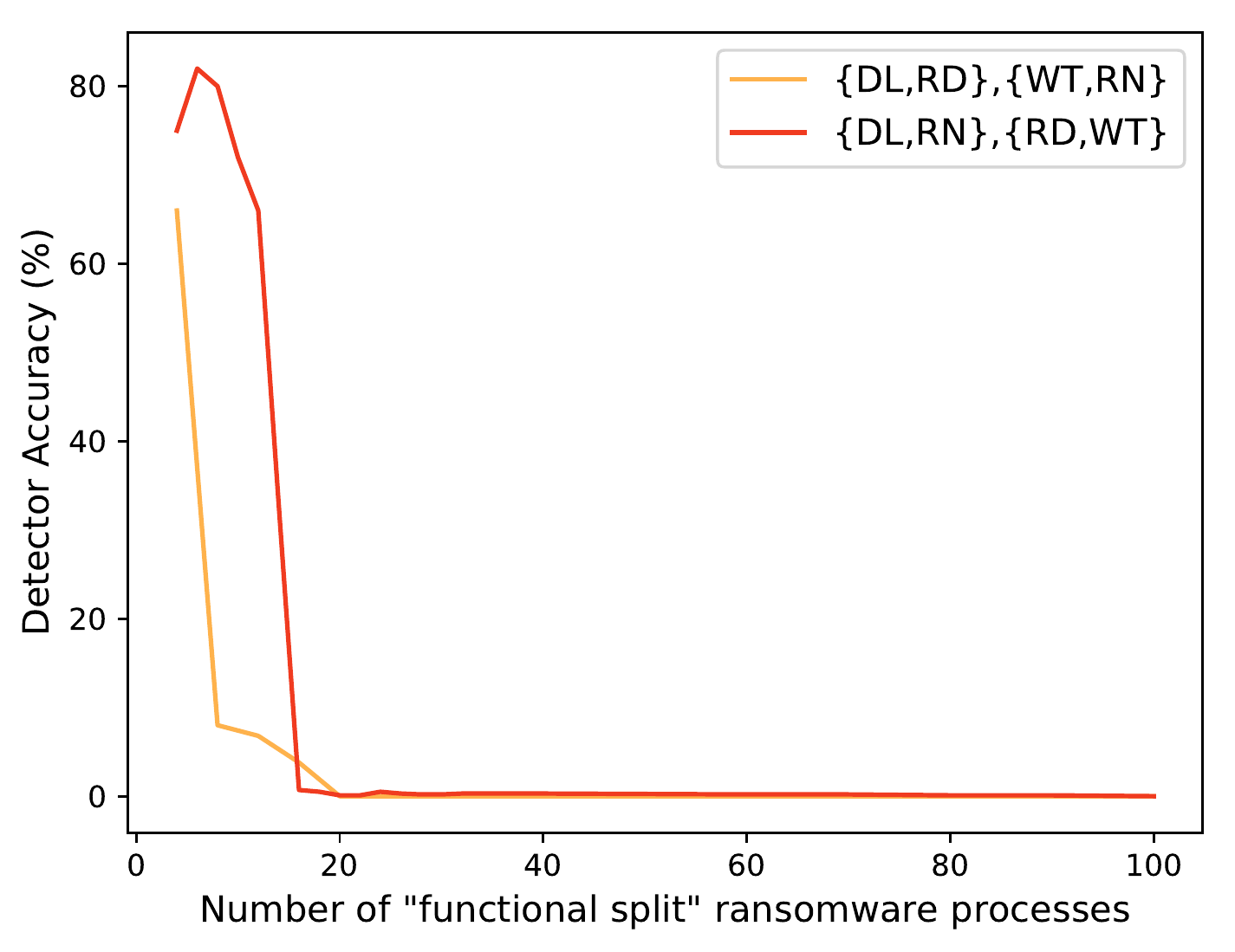}
            \vspace{-0.1in}
            \caption{Combined Functional Splitting}
            \label{fig:combined_functional_splitting}
        \end{subfigure}
        \caption{Evaluation of the functional splitting evasion technique against ShieldFS.}
        \vspace{-0.1in}
	\end{figure*}

First we evaluate single functional splitting, where each functional split process performs only one type of operation, resulting in four functional groups (DL, RD, WT and RN process groups). Within each functional group, we apply our process splitting technique: starting with a single functional split process per group, we recursively split it into multiple processes until complete evasion is achieved. As illustrated in Figure~\ref{fig:single_func_split}, we are able to completely evade ShieldFS by using 20 functional split processes, $5$ for each of the four functional groups. Note the contrast between Figure~\ref{fig:single_func_split} and Figure~\ref{fig:binary_splitting_eval_sfs}. With single functional splitting, $4$ processes (one for each functional group) are enough to drop the detector accuracy down to $\sim 2.5\%$, compared to the $\sim 7500$ processes required with process splitting. 

The effectiveness of functional splitting can be explained by analyzing the dataset. There is a significant difference in behavior, in terms of types of operations performed, between benign and ransomware processes over their lifetime. All of the ransomware processes in the dataset perform DL, RD, WT and RN types of operations, while only approximately $19\%$ of benign processes have a similar behavior. Since the feature expression profile between traditional and functional split ransomware is so different, with the latter being closer to benign processes than traditional ransomware, the accuracy of the classifier is heavily affected. To validate this hypothesis, we further study how different functional groups affect the performance of the detector. In particular, using combined functional groups (i.e. processes performing RD and WT, or DL and RN), rather than single functional groups, should result in higher detection accuracy as the behavioral profile of the functional split ransomware gets closer to that of a traditional ransomware. Figure~\ref{fig:combined_functional_splitting} illustrates our results.
This experiment evaluates the accuracy of ShieldFS considering two different implementations of functional split ransomware. In the first implementation, the operations are divided into the two functional groups  \{DL,RD\},\{WT,RN\}, while in the second implementation the two functional groups are \{DL,RN\},\{RD,WT\}. Figure~\ref{fig:combined_functional_splitting} shows that the initial accuracy of the classifier is much higher when compared to single functional splitting, hovering around $80\%$ for \{DL,RN\},\{RD,WT\} and around $70\%$ for \{DL,RD\},\{WT,RN\}. However, the accuracy quickly drops as we apply process splitting within each functional group, reaching $\sim 0\%$ at $20$ processes ($10$ for each functional group). The high initial detection accuracy for Figure~\ref{fig:combined_functional_splitting} is due to the fact that in the first ransomware implementation we have the \{RD,WT\} functional group and in the second implementation we have the \{WT,RN\} functional group. Both these functional groups are always present in traditional ransomware, therefore the model is more likely to classify processes that heavily exhibit these features as malicious. 
Indeed, we can see that after process splitting is applied in each functional group---and therefore the number of operations per functional split ransomware process decreases---the accuracy for both functional splitting implementations quickly falls towards zero.

\bparagraph{- Mimicry:}
Functional splitting evades detection by preventing the expression of one or more features from ransomware processes. Mimicry takes this concept one step further: rather than completely removing a feature, we model ransomware features so that, on average, they are identical to those of benign processes. We build our model of a typical benign process by performing an in-depth statistical analysis on the behavior of benign processes in the ShieldFS dataset~\cite{continella_shieldfs:_2016}, which comprises observations of well above one month of data from 2245 unique benign applications and $\sim1.7$ billion IRPs. We compute the average value for the main features used to profile ransomware and we extract the ratios between different types of I/O operations performed by benign processes. Finally, we split the ransomware activity into multiple processes, based on average feature values and ratios. We evaluate our approach on the ShieldFS ransomware traces that are part of our testing set.

We focus on modeling the four main operations performed by ransomware and benign processes---DL, RD, WT, RN---together with the number of file accessed. Note that we could easily consider more features in our modeling, up to all features described in Section~\ref{sec:Features}. However, since the goal of this evaluation is to prove the effectiveness of our techniques, it is sufficient to consider the most representative features. Table~\ref{tab:benign_process_behaviour} shows the different behavioral profiles exhibited by benign process, along with how represented that behavior is in the dataset. As can be seen, the most represented functional group of benign processes exhibits all four main operations \{DL,RD,WT,RN\}, with the functional groups \{DL,RD\} and \{RD,WT,RN\} being a close second and third. On the other hand, if we consider the behavioral profile of ransomware processes, all 383 ransomware samples perform all four main operations. Given that the first three process behavior groups in Table~\ref{tab:benign_process_behaviour} are all highly represented, any of them would be a suitable target for mimicry. For this evaluation, we decided to use the \{DL,RD,WT,RN\} functional group. While this functional group is also representative of most benign processes, the average number and ratio of operations is completely different when compared to ransomware. This functional group seems to be the worst-case scenario for our mimicry evasion technique. 
As illustrated in Table~\ref{tab:operation_ratio}, for benign processes in the \{DL,RD,WT,RN\} group, the ratio between operations is 1:16:13:1. This means that for each DL operation, there are $16$ RD, $13$ WT and $1$ RN operations respectively. Moreover, processes in this functional group access on average about $0.83\%$ of the total number of user files in the system. We split our ransomware traces in the test set by following these averages and ratios, resulting in $170$ mimicry ransomware processes, and successively query the classifier with each of them. We replicate this experiment for each of the $42$ ransomware sample in our test set. None of the mimicry processes for any of the $42$ ransomware are detected by the ShieldFS classifier.

\bparagraph{- Discussion:} It is worth noting the huge improvement gained with mimicry with respect to process splitting. In both mimicry and process splitting, each process performs all ransomware operations and therefore exhibits all the features used by ShieldFS for classification. However, with $170$ process-split ransomware the detection rate of ShieldFS is about $14\%$, while with mimicry we are able to fully evade the detector. In comparison, process splitting needs almost two orders of magnitude more processes to achieve full evasion ($11000$).

\smallsection{2) RWGuard}
This section evaluates the effectiveness of our techniques against the RWGuard ransomware detector.

\begin{figure}[t]
   \centering
   \includegraphics[width=0.9\columnwidth]{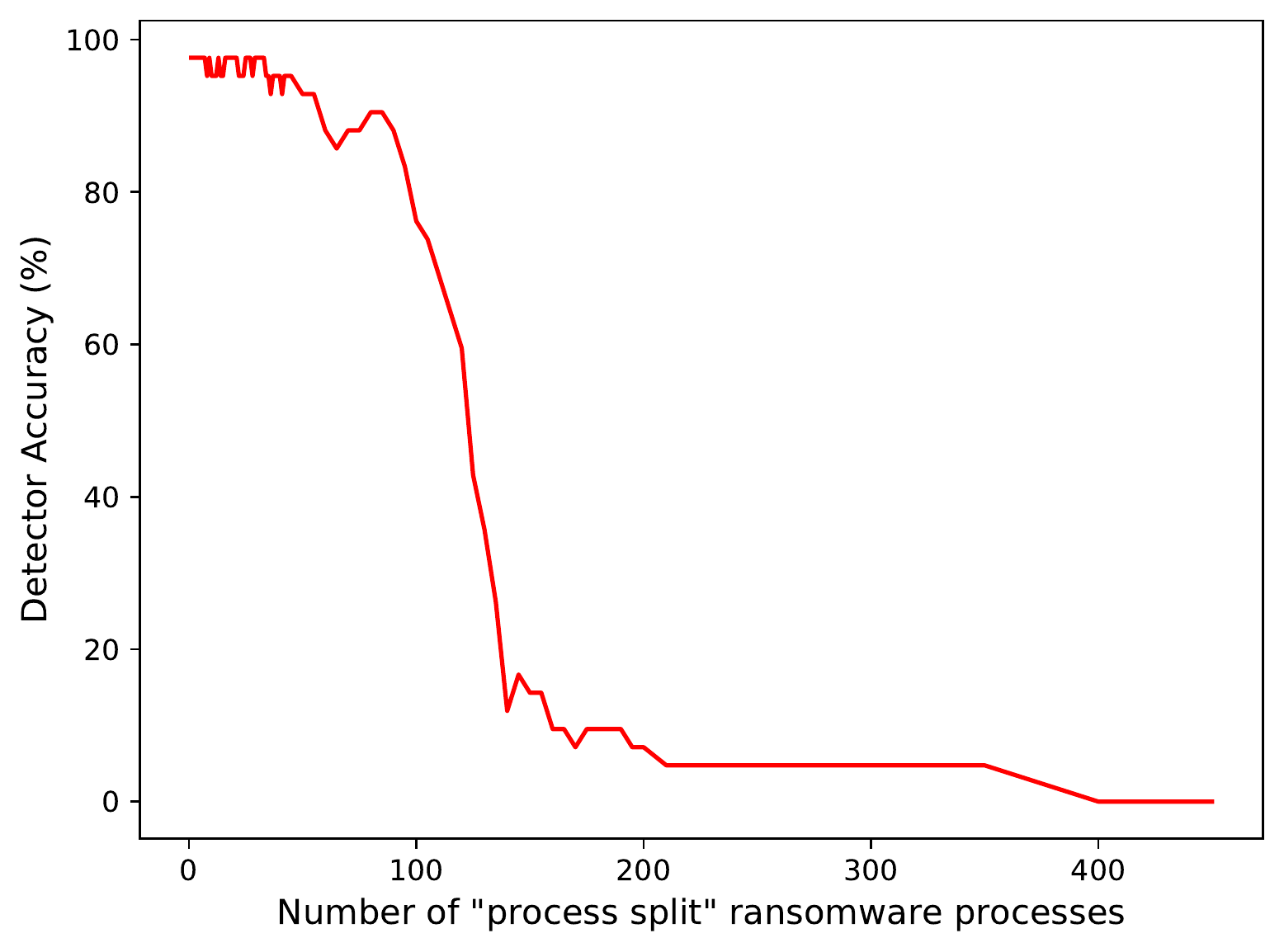}
   \caption{Evaluation of process splitting against RWGuard}
   \label{fig:rwguard_binary_splitting}
   \vspace{-0.1in}
 \end{figure}

\bparagraph{- Process Splitting:}
We implement process splitting as in the evaluation against ShieldFS. As illustrated in Figure~\ref{fig:rwguard_binary_splitting}, the detection accuracy for RWGuard follows a curve similar to that of ShieldFS: the accuracy of the classifier initially remains stable around the original $99.4\%$, until a critical point, after which it quickly decreases to $\sim 10\%$. Afterwards, both curves exhibit a long tail, with the detection accuracy very slowly decreasing to zero after $400$ processes for RWGuard.

\begin{figure*}[t]
	    \centering
	    \begin{subfigure}{.9\columnwidth}
	        \centering
            \includegraphics[width=.9\columnwidth]{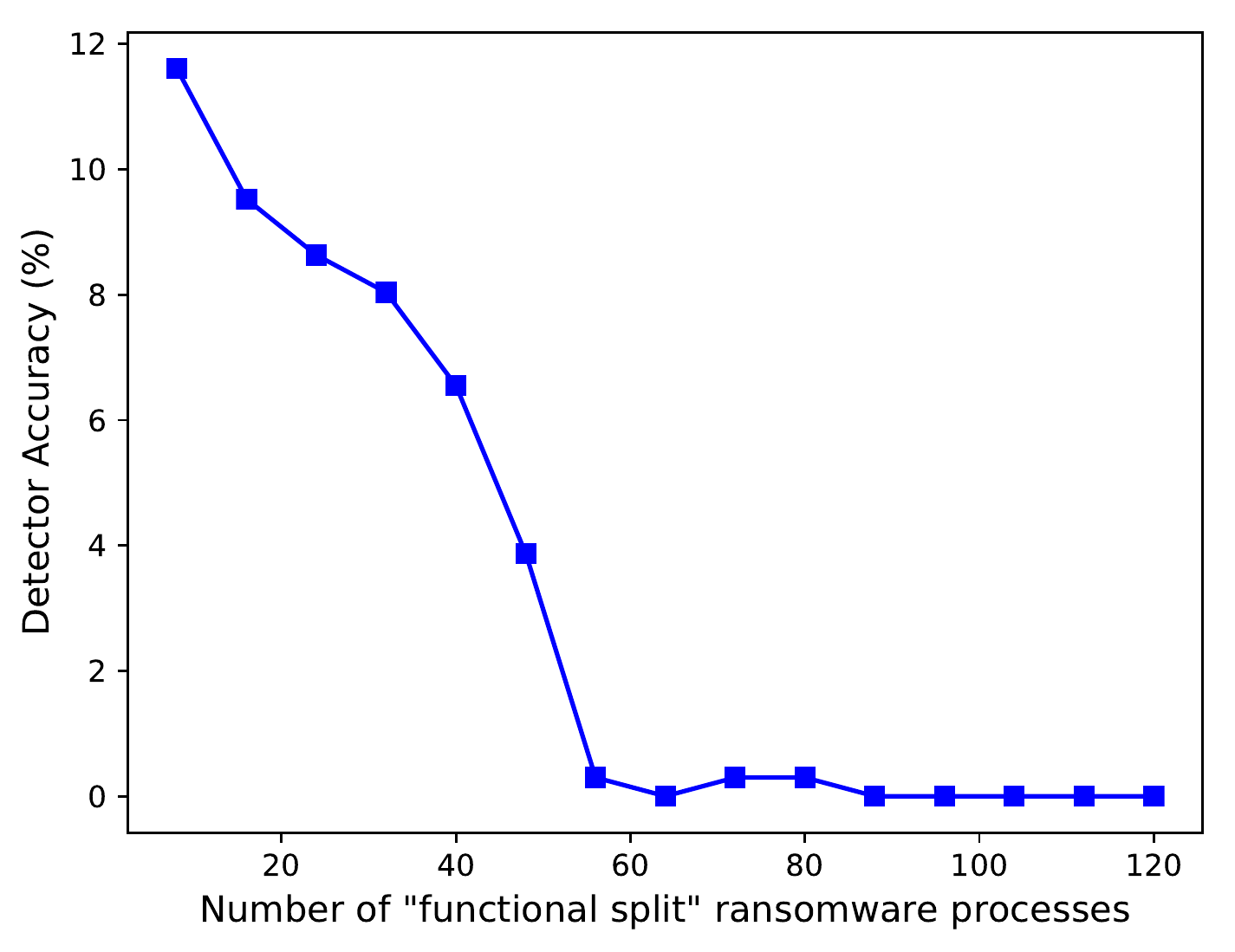}\
            \vspace{-0.1in}
            \caption{Single Functional Splitting}
            \label{fig:rwguard_single_splitting}
        \end{subfigure}
        \begin{subfigure}{.9\columnwidth}
            \centering
            \includegraphics[width=.9\columnwidth]{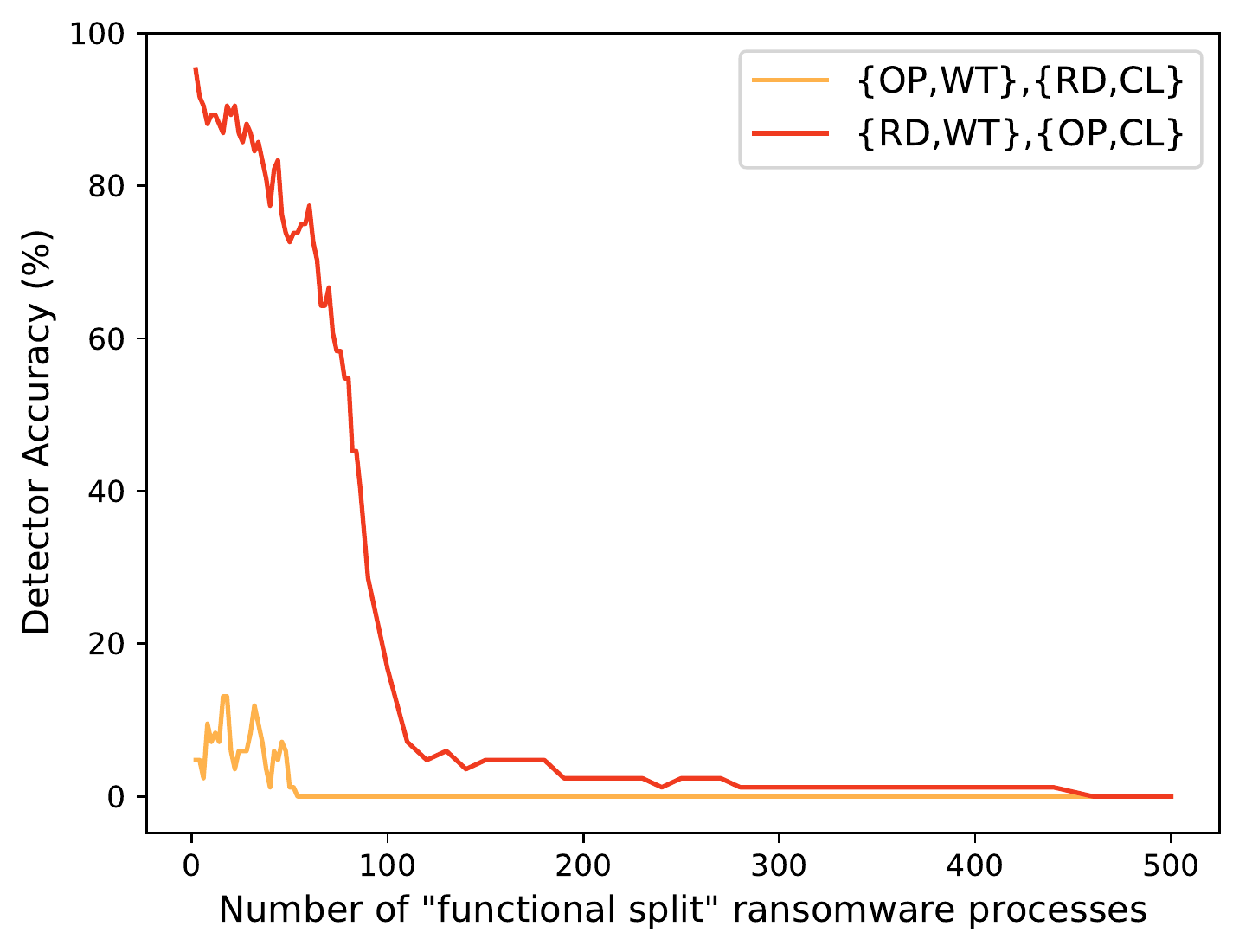}
            \vspace{-0.1in}
            \caption{Combined Functional Splitting.}
            \label{fig:rwguard_combined_splitting}
        \end{subfigure}
        \caption{Evaluation of the functional splitting evasion technique against RWGuard.}
        \vspace{-0.1in}
	\end{figure*}

\bparagraph{- Functional Splitting:}
The RWGuard detector uses eight features to classify benign and malicious processes: RD, WT, OP, CL, FRD, FWT, FOP and FCL. In this evaluation, we split the ransomware traces based on all eight features, and assess how each functional split ransomware process performs against the detector. We begin the evaluation with single functional splitting, where each functional split process performs only one type of operation, resulting in eight functional groups (one for each feature). Within each functional group, we apply process splitting until complete evasion is achieved. As shown in Figure~\ref{fig:rwguard_single_splitting}, to fully evade the RWGuard classifier we need $64$ functional split processes -- $8$ for each functional group.

        We further study how different functional groupings affect the accuracy of RWGuard. In particular, we evaluate the accuracy of RWGuard against two different implementations of functional split. In the first, the operations are divided into the two functional groups \{OP,WT\},\{RD,CL\}. For the second implementation, we use the \{RD,WT\},\{OP,CL\} functional groups. For the purpose of grouping, we make no distinction between normal and fast operations in this experiment. 
As shown in Figure~\ref{fig:rwguard_combined_splitting} and consistently with our ShieldFS evaluation (Figure~\ref{fig:combined_functional_splitting}), we see that the initial accuracy for \{RD,WT\},\{OP,CL\} is much higher than in the single functional splitting case, starting at approximately $95\%$ for two processes (one per functional group). This behavior is to be expected since RD and WT, two of the features with the highest importance for both detectors, are performed in the same functional group. Indeed, when we split these operations in two separate functional groups the accuracy of RWGuard is much lower, starting at $\sim 4\%$ with only $2$ processes (\{OP,WT\},\{RD,CL\} in Figure~\ref{fig:rwguard_combined_splitting}).

\bparagraph{- Mimicry:}
We evaluate our mimicry approach against the RWGuard classifier. As for the ShieldFS evaluation, we model ransomware features so they are, on average, identical to those of benign processes. In particular, we model the main features used by RWGuard: RD, WT, OP, CL, FRD, FWT, FOP and FCL. 
We split the ransomware traces in the test set by following the average operation number and operation ratio performed by benign processes, which resulted in 10 mimicry ransomware processes, and queried the classifier with each individual split trace. None of the 42 ransomware samples in our test set were detected by RWGuard.

\subsection{Cerberus Evaluation}
\label{sec:CerberusEvaluation}
This section evaluates Cerberus, our ransomware prototype implementing functional splitting and mimicry evasion. Section~\ref{sec:LogEvaluation} showed that our evasion techniques are effective when applied on the traces generated by traditional ransomware. However, it is still necessary to demonstrate that a prototype implementation of our techniques would work in practice. Moreover, in Section~\ref{sec:LogEvaluation} we modeled mimicry processes based on the benign processes of the dataset used for the training of ShieldFS and RWGuard. It is necessary to determine whether our techniques can generalize to the case where the benign process model is derived from a surrogate dataset (i.e. a dataset different from the one used to train the classifier).

\begin{table}[t]
\centering
\scriptsize
\begin{tabular}{|l|l|l|l|l|l|l|l|}
\hline
\multicolumn{1}{|c|}{\textbf{Combination}} & \multicolumn{1}{c|}{\textbf{DL}} & \multicolumn{1}{c|}{\textbf{RD}} & \multicolumn{1}{c|}{\textbf{WT}} & \multicolumn{1}{c|}{\textbf{RN}} & \multicolumn{1}{c|}{\textbf{\begin{tabular}[c]{@{}c@{}}RD\\ Entropy\end{tabular}}} & \multicolumn{1}{c|}{\textbf{\begin{tabular}[c]{@{}c@{}}WT\\ Entropy\end{tabular}}} & \multicolumn{1}{c|}{\textbf{\begin{tabular}[c]{@{}c@{}}File\\ Access\end{tabular}}} \\ \hline
RD,RN       & 0     & 2     & 0     & 1     & 0.53                       & 0                          & 0.02\%                             \\\hline
WT          & 0     & 0     & 1     & 0     & 0                          & 0.42                       & 0.60\%                             \\\hline
DL,RD,WT,RN & 1     & 16    & 13    & 1     & 0.59                       & 0.46                       & 0.83\%                             \\\hline
RD          & 0     & 1     & 0     & 0     & 0.46                       & 0                          & 0.03\%                             \\\hline
WT,RN       & 0     & 0     & 5     & 1     & 0                          & 0.47                       & 0.02\%                             \\\hline
RD,WT       & 0     & 5     & 1     & 0     & 0.29                       & 0.57                       & 1.33\%                             \\\hline
DL,RD,RN    & 8     & 39    & 0     & 1     & 0.42                       & 0                          & 0.09\%                             \\\hline
DL,WT       & 2     & 0     & 1     & 0     & 0                          & 0.51                       & 0.01\%                             \\\hline
RD,WT,RN    & 0     & 6     & 20    & 1     & 0.53                       & 0.28                       & 0.22\%                             \\\hline
DL,RD,WT    & 3     & 52    & 1     & 0     & 0.57                       & 0.77                       & 0.17\%                             \\\hline
DL          & 1     & 0     & 0     & 0     & 0                          & 0                          & 0.00\%                             \\\hline
DL,RD       & 1     & 2     & 0     & 0     & 0.52                       & 0                          & 0.17\%                             \\\hline
DL,WT,RN    & 1     & 0     & 8     & 2     & 0                          & 0.39                       & 0.03\%                             \\\hline
DL,RN       & 45    & 0     & 0     & 1     & 0                          & 0                          & 0.06\%                             \\\hline
RN          & 0     & 0     & 0     & 1     & 0                          & 0                          & 0.03\%                             \\\hline
\end{tabular}
\caption{Ratio between different operations for various types of benign processes.}
\label{tab:operation_ratio}
\vspace{-0.15in}
\end{table}

\smallsection{1) ShieldFS}
We evaluate Cerberus against ShieldFS in our virtual machine, both in the functional split and mimicry modes. Cerberus implements functional splitting with the following three functional groups: \{DL\},\{WT\},\{RD,RN\}. Reading and renaming operations are performed by the same process group mainly for implementation convenience. It is worth noting that aggregating these two features makes it easier for the classifier to detect the ransomware. 
By setting Cerberus to use $6$ processes per functional group ($18$ processes total, which is the closest to the $20$ processes suggested by our trace-based evaluation), we were able to fully evade the detector: no functional split process was flagged as ransomware.

We also evaluate Cerberus in mimicry mode against ShieldFS. We described the details of the mimicry implementation in Cerberus in Section~\ref{sec:Cerberus}. The number of processes in mimicry mode depends on the average number of files accessed by the mimicked benign process group in our dataset. Table~\ref{tab:operation_ratio} shows that \{DL,RD\} processes access on average $\sim 0.17\%$ of the total files, while \{RD,WT,RN\} processes access $\sim 0.22\%$. In our VM, this results in a Cerberus run with $470$ mimicry processes, which were all able to evade the ShieldFS detector, fully encrypting the VM files. This evaluation proves that our attacks are practical and applicable in realistic settings.

\smallsection{2) RWGuard}
We further evaluate Cerberus against RWGuard in our virtual machine, both in the functional split and mimicry modes. An important difference compared to the RWGuard evaluation in Section~\ref{sec:LogEvaluation} is that functional splitting in Cerberus considers only three functional group, that is: \{DL\}, \{WT\} and \{RD,RN\}. Cerberus does not split RWGuard-specific features (i.e., OP, CL, FOP, FCL, FRD, FWT). Regardless of this fact, we are able to fully evade RWGuard with Cerberus set to use $18$ functional split processes in total ($6$ per functional group), as in the ShieldFS case.

We also evaluate the mimicry mode of Cerberus against RWGuard. For this evaluation, Cerberus is trained with the model of benign processes obtained from the ShieldFS dataset, while the RWGuard model is trained on the original dataset used by the authors in~\cite{mehnaz_rwguard}. As before, in our VM evaluation Cerberus runs with $470$ mimicry ransomware processes, which are all able to fully evade the RWGuard detector, fully encrypting the VM files. This evaluation shows that our evasion techniques can generalize to classifiers trained on different datasets.

\subsection{Evaluation against a commercial detector}
\label{sec:CommercialDetectorEval}
In previous experiments, the features used by the detectors (ShieldFS and RWGuard) were known. However, in a real attack scenario this white-box setting assumption might not hold true. The last part of our experimental evaluation focuses on black-box settings where details of the detectors are not known. In particular, we pitch Cerberus against a leading commercial ransomware detector: \textbf{[anonymized]} . \textbf{[anonymized]} leverages machine learning techniques to detect the presence of ransomware in the system. We have no knowledge of the internal workings of the \textbf{[anonymized]} classifier, such as which features it uses for classification, nor of its dataset. This makes \textbf{[anonymized]} an ideal detector to test the viability of our evasion techniques in a black-box setting. We evaluate Cerberus against \textbf{[anonymized]} in both the functional split and mimicry modes. For the functional splitting approach, we continue to set Cerberus to use a total of $18$ functional split processes ($6$ per functional group). All $18$ functional split processes successfully evade \textbf{[anonymized]}, fully encrypting the VM files.

We also evaluate Cerberus running in mimicry mode against \textbf{[anonymized]} . As usual, the mimicry behavior of Cerberus processes is modeled based on the ShieldFS benign process dataset. Therefore, Cerberus runs with the usual $470$ mimicry ransomware processes, which all successfully evade \textbf{[anonymized]} and fully encrypt the VM files. This last experiment shows that our evasion techniques are general, are effective on commercial detectors and work in a black-box setting where we have no information on the classifier.


\section{Discussion}
\label{sec:Discussion}

\paragraph{Countermeasures} It is in principle possible to dynamically track process-to-process communication to identify clusters of cooperating processes. A detector may then compute behavioral features over the cluster of processes. This is however challenging, since processes can communicate---both for legitimate and malicious reasons---in a variety of manners which must all be tracked (e.g., OS pipes, shared memory, file system). 
Tracking process-to-process communication reliably and without generating undue overhead is not straightforward. Moreover, this complexity is further exacerbated by the possibility of using covert channel techniques, rather than standard inter-process communication functions. Devising a way to reliably identify a set of communicating processes is an interesting direction for future work. 

Another approach entails identifying \textit{synchronized process behavior} across applications running concurrently in different machines. This approach leverages the insight that a ransomware infection typically involves an entire network. Similar approaches, although based on network traffic, have proven effective for botnet nodes detection~\cite{gu_bothunter:_2007}. We note that both the functional splitting and mimicry attack can, by design, split operations in arbitrarily different ways. This enables randomizing the attack behavior across different machines.

Finally, many defenses against adversarial attacks---both theoretical and practical---have been proposed (e.g.,~\cite{tong_hardening_2017, carlini_adversarial_2017}). We posit that the current generation of behavioral malware detectors exhibits \textit{feature vulnerability}~\cite{maiorca2017adversarial}, i.e., it is possible to generate malicious behavior that looks, feature-wise, exactly like benign behavior. This suggests that increasing the sophistication of the classifier without rethinking the features, may not suffice to remediate such attacks.

\paragraph{Generalization} While we concretize our attack in the context of ransomware, we believe the underlying principle has general applicability. Behavioral malware detectors tend to leverage system calls or their effect (e.g., IRPs in Windows) issued by individual processes in the course of their functioning. Detection algorithms look at sequence of calls (e.g., RNN-based approaches~\cite{pascanu_malware_2015}), call frequency/count~\cite{continella_shieldfs:_2016, mehnaz_rwguard, kirda_redemption}, or more complex inter-call data dependencies~\cite{fredrikson_synthesizing_2010}. In principle, any of the approaches above can be fooled by partitioning system calls across different processes, thus avoiding generation of sequences with the expected properties. A complication is that calls may depend on previously established process state; e.g. a file can only be read if it has previously been opened by the same process. However, many types of state can be communicated across processes (e.g., using {\small \tt DuplicateHandle()} in Windows). Therefore, attackers have significant freedom in partitioning control flow of malicious programs across processes.

\vspace{-0.02in}
\section{Related Work}
\label{sec:RelatedWork}

\paragraph{Ransomware detection} For a review of behavioral ransomware detection techniques~\cite{continella_shieldfs:_2016, mehnaz_rwguard, kirda_redemption, amin_kharraz_unveil:_2016, scaife_cryptolock_2016}, the reader is referred to Section~\ref{sec:BehavioralRansomwareDetection}.
Other proposals focus specifically on randomness of written data to identify encrypted content. Data-aware Defense~\cite{lipmaa_data_2017} performs the $\chi^2$-test on a sliding window of write operations. Mbol et al.~\cite{foresti_efficient_2016} use a test based on the Kullback-Liebler divergence to detect ransomware converting high-entropy JPEG files to encrypted content. Depending exclusively on randomness is dangerous for reasons pointed out in Section~\ref{sec:WriteEntropy}.
An orthogonal line of work proposes the use of \textit{decoy files} for ransomware detection~\cite{genc_deception_2019, moore_detecting_2016, moussaileb_ransomwares_2018}. Such defenses are outside the scope of our work. We believe decoys are a promising strategy, but they raise usability concerns, and their evasion has been poorly studied.

\paragraph{Multiprocessing in existing malware} Several existing ransomware families use multi-processing.
This happens for example in WannaCry. Encryption is still performed by one process, while the others perform non encryption-related auxiliary tasks~\cite{wannacry_analysis}. Petya/NotPetya works in a similar way~\cite{petya_analysis}. The CERBER ransomware (not to be confused with our \textit{Cerberus} prototype), despite its name, does not appear to perform multi-process encryption. While it has been claimed that CERBER attempts to evade machine learning, these claims refer to obfuscation of static payload features~\cite{cerber_analysis}.

\paragraph{Evasion of ransomware detectors} The work closest in spirit to ours is the critical analysis of ransomware defenses by Genç et al.~\cite{gruschka_next_2018}. For what concerns behavioral detection, their work is more limited in scope than ours and consider a smaller set of features. Furthermore, the work by Genç et al does not incorporate the notion of mimicry and only focus on simple feature obfuscation (e.g., avoid changing file types).

\paragraph{Adversarial machine learning} There is a line of work in adversarial machine learning that focuses on generating adversarial samples for various classes of malicious programs. In the mobile malware domain, Grosse et al.~\cite{10.1007/978-3-319-66399-9_4} generate malicious Android app packages which go undetected by a custom neural network classifier which uses manifest-derived features. This attack uses Papernot's method~\cite{papernot2016limitations} to guide mutations to app manifests.

Other attacks are not specifically focused on mobile. Anderson et al.~\cite{anderson_evading_2017} propose the use of reinforcement learning to guide mutations to malicious executables to make them undetectable. Rosenberg et al.~\cite{rosenberg_generic_2018} generate adversarial malware binaries by altering various static features using a custom mimicry attack. Hu and Tan~\cite{hu_generating_2017} propose the use of generative adversarial networks (GANs) to mutate feature vectors derived from the presence/absence of imported DLLs and API calls in a malware binary. \cite{hu_generating_2017} does not propose a method to concretely generate malware binaries, only feature vectors.

Finally, several works propose attacks which mutate the static structure of PDF-based exploits to prevent their detection~\cite{hutchison_evasion_2013, Srndic:2014:PEL:2650286.2650798, xu2016automatically, dang_evading_2017}.

All the works above focus on static features, i.e., they alter the appearance of a malicious file object, but not its run-time behavior.

There is limited work on attacking dynamic (behavioral) features---i.e., features generated by actions performed by a process at run-time. Arguably, such features are harder to attack; indeed, they represent actions that a malware needs to execute in order to achieve its malicious goals.

Rosemberg et al.~\cite{rosenberg_generic_2018, rosenberg_query-efficient_2018} and Hu and Tan~\cite{hu_black-box_2017} proposed methods to defeat malware detectors trained on dynamically-generated sequences of API calls. These proposals work by perturbing the sequence of API calls, chiefly by inserting dummy calls.

 While we use dummy calls as part of our mimicry attack, we also leverage a broader set of capabilities such as distributing calls across processes to reduce feature expression. This give our technique the ability to decrease per-process frequencies/counts of certain calls without slowing down the attack (necessary to defeat the detectors in our evaluation), or to obfuscate data dependencies between calls (such dependencies are used by some detectors, e.g.~\cite{fredrikson_synthesizing_2010}).

\vspace{-0.02in}
\section{Ethical Considerations}
\label{sec:Ethical}

 Our work is motivated by the interest in understanding the limitations of the current generation of malware detectors. We believe that doing so is necessary to ensure that detection algorithms evolve and remain effective against the constantly-shifting threat landscape. We \textbf{do not} plan to publicly release our ransomware prototype (Section~\ref{sec:Cerberus}), in order to prevent its use in threat development. However, we do plan to make it available on a case-by-case basis to reputable research groups. Furthermore, prior to submitting this paper we communicated details and results of our attack to \textbf{[anonymized]}.


\vspace{-0.02in}
\section{Conclusions}
\label{sec:Conclusions}

We demonstrated a novel practical attack against behavioral malware detectors. Our attack splits malware operations across a set of cooperating processes. This is done in such a way that no individual process behavior is flagged as suspicious by a behavioral process classifier, but the combined behavior successfully accomplishes a malicious goal.

We concretely defined and implemented this concept in the ransomware domain. We proposed three novel attacks, \textit{process splitting}, \textit{functional splitting}, and \textit{mimicry}. Evaluation shows that our methods successfully evade state-of-the-art detectors without limiting the capabilities of ransomware. 

To the best of our knowledge, this is the first comprehensive evaluation of this attack model in the domain of malware and ransomware in particular. Our work individuates a significant limitation in behavioral detection algorithms and has relevant practical implications for the current generation of malware detectors. 
Directions for future work include demonstrating concrete applications to other malware domains, and devising efficient detection algorithms that are robust to our attacks.

\bibliographystyle{IEEEtran_ldk}

\bibliography{bibliography}

\end{document}